\documentclass[prl,twocolumn,preprintnumbers,amsmath,amssymb,superscriptaddress]{revtex4}

\usepackage{ulem}
\usepackage[T1]{fontenc}
\usepackage[utf8]{inputenc}
\usepackage[german,english]{babel}
\usepackage{epsfig}
\usepackage{amssymb}
\usepackage{amsmath}
\usepackage{braket}
\usepackage{bm}
\usepackage{color}
\usepackage{times}
\usepackage[colorlinks,bookmarks=false,citecolor=blue,linkcolor=red,urlcolor=blue]{hyperref}
\usepackage{graphicx}
\usepackage{soul}

\newcommand{\beq}{\begin{equation}}
\newcommand{\eeq}{\end{equation}}
\newcommand{\bea}{\begin{eqnarray}}
\newcommand{\eea}{\end{eqnarray}}
\newcommand{\nn}{\nonumber}

\def\lsi{\raise0.3ex\hbox{$<$\kern-0.75em\raise-1.1ex\hbox{$\sim$}}}
\def\gsi{\raise0.3ex\hbox{$>$\kern-0.75em\raise-1.1ex\hbox{$\sim$}}}

\begin{document}
\setstcolor{red}
\title{
Zero-point excitation of a circularly moving detector in an atomic condensate \\ and phonon laser dynamical instabilities 

}

\author{Jamir Marino} 
\thanks{These two authors contributed equally}
\affiliation{Department of Physics, Harvard University, Cambridge MA 02138, USA\\
Department of Quantum Matter Physics, University of Geneva, 1211, Geneve, Switzerland\\
Institut f\"ur Physik, Johannes Gutenberg Universit\"at Mainz, D-55099 Mainz, Germany}

\author{Gabriel Menezes} 
\thanks{These two authors contributed equally}
\affiliation{Department of Physics, University of Massachusetts Amherst, MA 01003, USA\\
Departamento de Fisica, Universidade Federal Rural do Rio de Janeiro, 23897-000, Seropedica, RJ, Brazil
}

\author{Iacopo Carusotto} 
\affiliation{INO-CNR BEC Center and Department of Physics, University of Trento, I-38123 Povo, Italy}



\begin{abstract} 
{We study  a circularly moving impurity in an atomic condensate for the realisation of superradiance phenomena in tabletop experiments. The impurity is coupled to the density fluctuations of the condensate and, in a quantum field theory language, it serves as an analog of a detector for the quantum phonon field. For sufficiently large rotation speeds, the zero-point fluctuations of the phonon field induce a sizeable excitation rate of the detector even when the condensate is initially at rest in its ground state. For spatially confined condensates and harmonic detectors, such a superradiant emission of sound waves provides a dynamical instability mechanism leading to a new concept of phonon lasing. Following an analogy with the theory of rotating black holes, our results suggest a promising avenue to quantum simulate basic interaction processes involving fast moving detectors in curved space-times.}

 \end{abstract}


\date{\today}
\maketitle


\textit{Introduction ---} Since Unruh's pioneering proposal in 1981~\cite{UnruhBH}, the last decades of research activities have witnessed the surge of  a new field where   concepts of general relativity and of quantum field theories in curved backgrounds are investigated in the so-called analog models of gravity~\cite{Living}. 
As a most celebrated example, acoustic analogs of black holes have been studied in trans-sonically flowing atomic Bose-Einstein condensates: the acoustic black hole horizon corresponds to the interface between regions of respectively sub- and super-sonic flow and was anticipated to emit a thermal radiation of phonons via Hawking processes~\cite{BHoles}. The first experimental observations of such phenomena~\cite{Jeff} were instrumental in triggering the on-going explosion of the field, with a revived interest in using analog models to investigate a variety of different effects of quantum field theories in curved space times, from the dynamical Casimir effect~\cite{Cas}, to acceleration radiation~\cite{Unruhacc}, to vacuum friction and Casimir forces~\cite{vac, Marino2017}. 

The subject of the present work is the phenomenon of \textit{rotational superradiance}~\cite{and,SR:book}, namely the amplification of classical waves reflected by a fast rotating body. In the simplest formulation, superradiance {processes are efficient} whenever the linear velocity of the object (or of parts of it) exceeds the phase velocity of the waves, so that the wave frequency seen in the comoving frame turns negative. Such negative-energy modes then provide the energy that is required to amplify positive-energy waves via superradiant effects.
In cylindrical geometries, a mode of frequency $\omega$ and azimuthal quantum number $n$ can be superradiantly amplified when the angular velocity $\Omega$ of the rotating body satisfies $\Omega>\omega/n$. 

Being a consequence of basic kinematical arguments, superradiance is an ubiquitous phenomenon in physics. Its first incarnation was the discovery of amplification of acoustic waves hinging  upon a supersonically moving boundary~\cite{andrei}, or  the amplification of  cylindrical electromagnetic waves {interacting with a rotating} material~\cite{zel}. In an astrophysical rotating black hole, the superradiant amplification of waves is a consequence of the space-like character of the generator of time translations inside the ergosphere~\cite{SR:book,GSM:17}, and it is at the root of the several instability phenomena of Kerr black holes~\cite{SU, inst}. {In the framework of analog models, theoretical studies have investigated superradiant phenomena in rotating classical and quantum systems~\cite{acous}, offering new interpretations of basic hydrodynamic phenomena~\cite{Giacomelli}. Experimental evidence of superradiant scattering of  {classical} surface waves on water was reported in~\cite{wein}. Much less studied are the quantum features when superradiant processes are triggered by zero-point fluctuations,  {and the even more intriguing quantum friction effects that result from back-reaction of superradiance on the rotational motion~\cite{Cal,backreaction}.}

%
%
%
\begin{figure}[t!]
\includegraphics[width=7.cm]{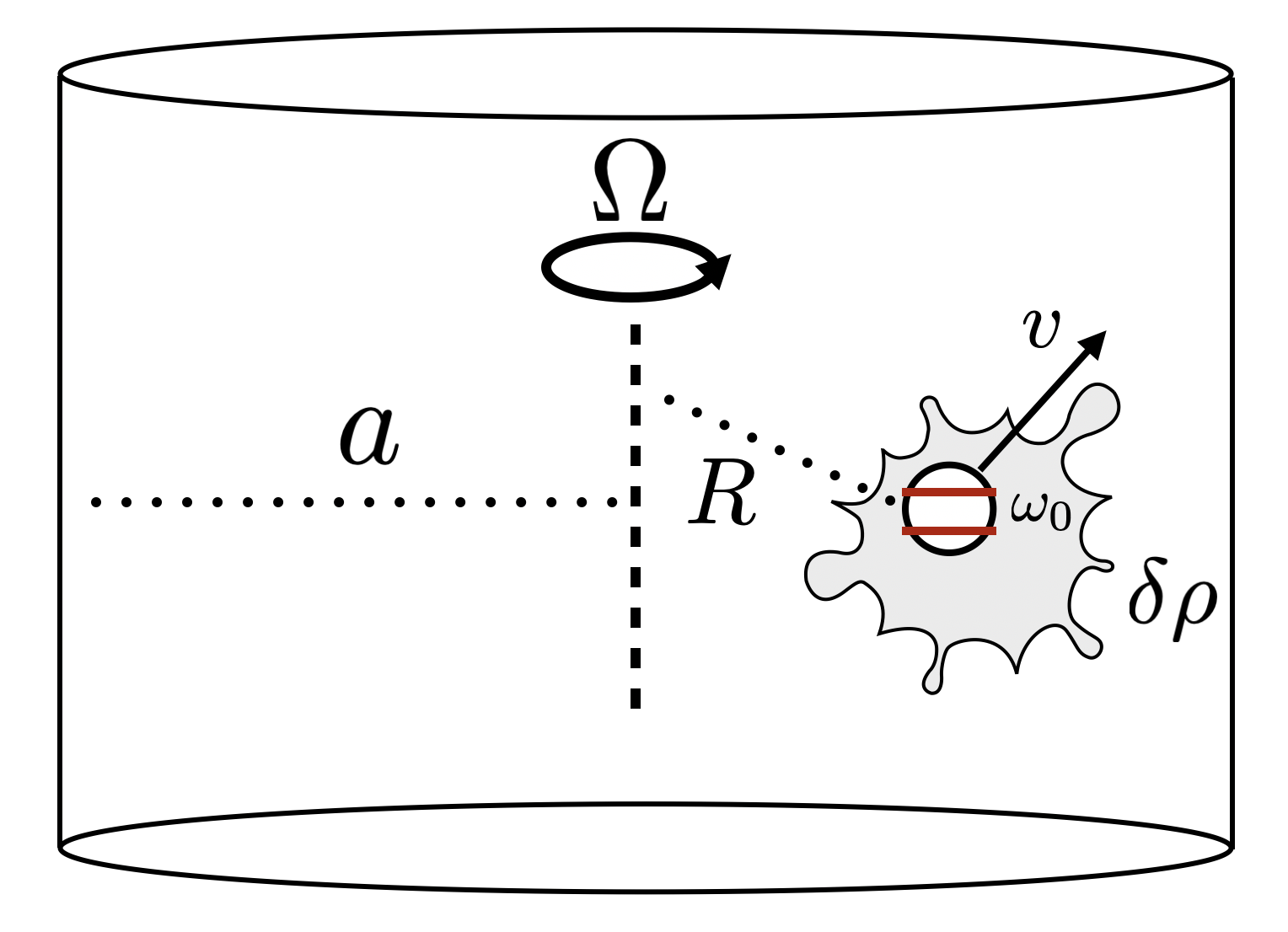}
 \caption{  A two-level detector with internal frequency, $\omega_0$, rotates with uniform angular velocity $\Omega$ at a distance $R$ from its rotational axis. The impurity couples to the density fluctuations, $\delta\rho (\mathbf{r})$, of a weakly interacting Bose gas, which is trapped inside a cylindrical cavity of radius $a$; the cavity is modeled via Dirichlet boundary conditions on the density fluctuations, $\delta\rho(\mathbf{r})|_{r=a}=0$. } 
\label{fig1}
\end{figure} 


In this {Rapid Communication}, we investigate novel superradiant  {phenomena} that  can occur in ultracold atomic systems. {In contrast to the conventional case of rotating fluids  considered in  superradiance, we} study a configuration where} the quantum fluid is at rest but {a neutral impurity} moves (classically)  
through the cold gas {at sufficiently large speeds}, as sketched in Fig.~\ref{fig1}.
{As it has been originally discussed in Refs.~\cite{Unruhacc,Marino2017}, the neutral impurity plays the role of a two-level detector in a canonical quantum field theory setup,
and it allows to explore superradiant phenomena beyond the usual amplification of incident waves~\cite{SR:book,acous}. Starting from the quantum vacuum state of the phonon field, we predict the spontaneous excitation} of the  internal degrees of freedom of the impurity in response to zero-point quantum fluctuations in the condensate{, which in turn get amplified into real phonons. In a trapped geometry, {the finite size of the fluid} provides a perfectly reflecting cavity for the phonon modes; in this way, the self-stimulation of the detector leads to a dynamical instability   for sound waves, which is rooted in superradiance. In contrast to the usual laser operation which requires an external pumping of the gain medium, the {{\it phonon lasing} mechanism} envisioned here is  {driven} by the mechanical motion of a detector that is initially prepared in its ground state. 

}



\textit{A circularly moving impurity in a uniform condensate ---} Consider a two-level {phonon} detector with internal frequency 
$\omega_0$ in circular motion with constant angular velocity $\Omega$ at distance $R$ from its  rotation axis (see Fig.~\ref{fig1} for an illustration). {As originally proposed in~\cite{Unruhacc,Marino2017},} the detector {is assumed to be coupled} to density fluctuations of a weakly interacting three dimensional Bose gas. Such a detector can be realized by means of an atomic quantum dot~\cite{atomicdot, fisher}, namely an impurity atom immersed in the condensate and collisionally coupled to the Bose gas via two  channels.
 {The first term is reminiscent of the interaction of a charged particle to an external scalar potential and can be cancelled via proper tuning of the interaction constants (e.g. via Feshbach resonances). In this case, only the second interaction term survives, and we find the hamiltonian} %
\begin{equation}\label{eq:hami}
H = \frac{\omega_0}{2}\sigma_{z}+ g_{-} \sigma_{x}\delta\rho({\bf r})+H_{\textit{B}},
\end{equation}%
where  {$\sigma_{x}, \sigma_{z}$ are quantum operators (proportional to the Pauli matrices) associated with the two-level detector, while $\delta\rho({\bf r})$ are the density fluctuations  of the BEC which couple to the detector  via the coupling constant $g_-$.} {It is immediate to recognize how this Hamiltonian closely resemble the dipole coupling between a  {neutral polarizable object} and the electromagnetic field.}  {Extension of the theory to detectors with a harmonic oscillator internal structure is straightforward and it will be discussed in the last part of the {Rapid Communication}.}


{Under  the standard weak interaction limit for  Bose gases,} density fluctuations {can be} treated within Bogolyubov theory~\cite{books}. The Bose gas Hamiltonian is, therefore, given by the usual expression
$H_{B} = \int d{\bf q}\, \omega_{{\bf q}}\,b^{\dagger}_{{\bf q}}b_{{\bf q}}$,
where $b^{\dagger}_{{\bf q}}, b_{{\bf q}}$ are the   creation and annihilation operators of  Bogolyubov quanta  {of momentum $\bf q$},  {whose dispersion relation is} $\omega_{{\bf q}}$.  
%
%
%
As customary, we refer with  $m$ to the mass of the condensate's particles, with $\mu$ to the condensate's chemical potential, with $c$ to its speed of sound, and with $\xi$ to its healing length.

%

%

 %

\textit{Vacuum excitation rate ---}  
Since the motion of the detector is non-inertial, we can expect a non-vanishing transition probability $A_{\uparrow}$ for the detector to jump from the ground to the excited state even for a condensate initially in its ground state. This effect is due to the {zero-point} quantum fluctuations in the phonon quantum vacuum and it {is associated with} the emission of a phonon. 
Related excitation mechanisms have been discussed for a detector in uniform super-sonic motion along a rectilinear trajectory in {standard electromagnetism, the so-called Ginzburg effect~\cite{Ginzburg}, as well as} in a BEC analog model of the latter~\cite{Marino2017}. 

While  {the emission from circularly moving detectors with relativistic accelerations} shares analogies with the Unruh effect~\cite{Birrell:82}, it is important to highlight a crucial difference. {{For circular motion, the detector} is not expected to emit radiation thermally equilibrated at the Unruh temperature~\cite{Bell:83,Bell:87,Holzmann:95,Unruh:98,Rad:12}, contrary to its {linearly accelerating} counterpart: attempts to define an effective temperature becomes problematic  in the spatial region $r>c_{l}/\Omega$ (where $c_{l}$ is the speed of light), since inconsistencies related to  causality prevent to define unambiguously a concept of a rotating vacuum and to build excitations on top of it~\cite{unr}.} 

\begin{figure}[t!]
\includegraphics[width=8.1cm]{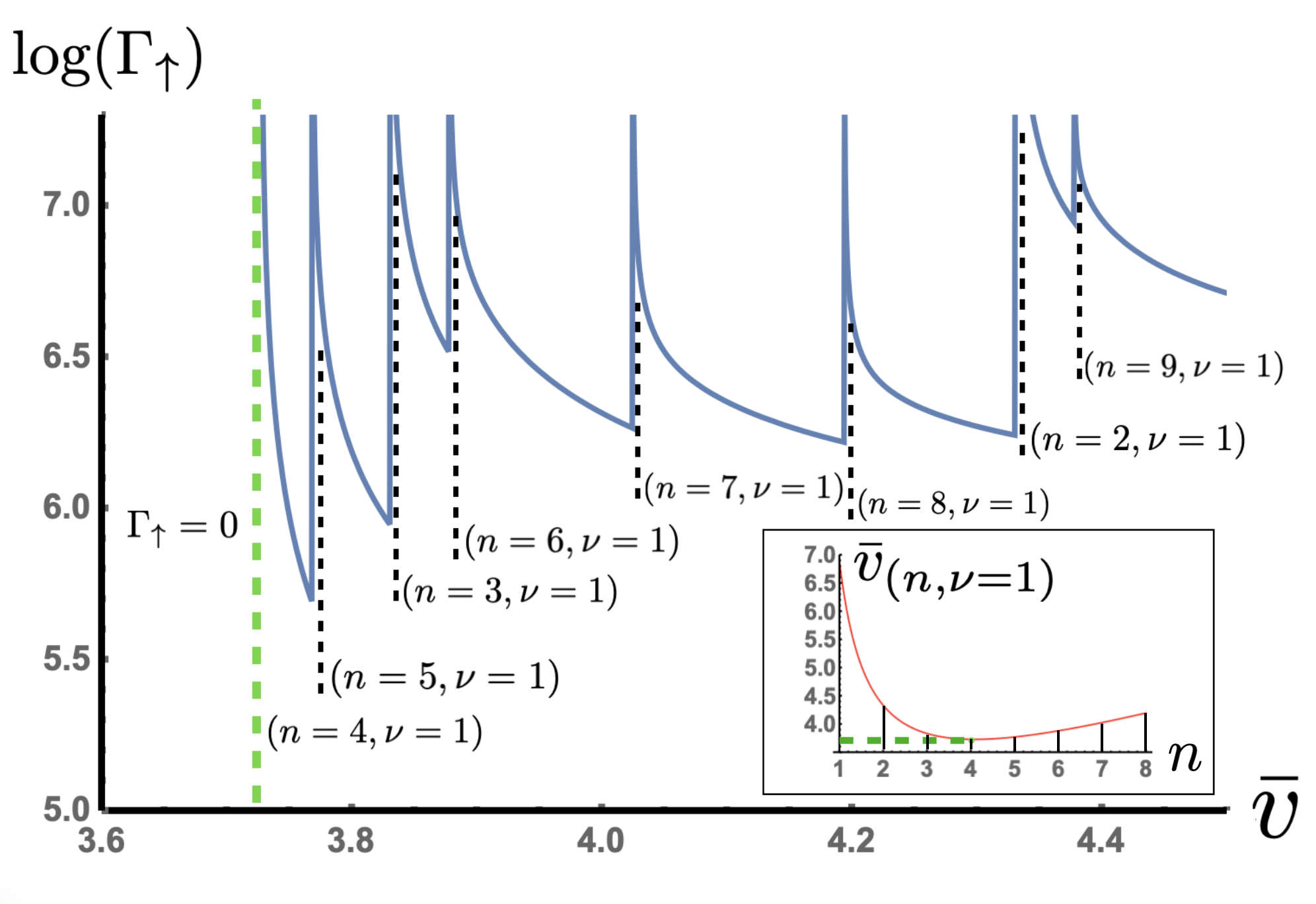}
 \caption{The dimensionless excitation rate, $\Gamma_{\uparrow}=A_\uparrow\,\omega_0/P(\omega_0,R)$ (with $P(\omega_{0}, R) = {g_{-}^2 \rho_{0}}/({2 m \omega_{0} R^{5} })$), evaluated for $\bar{y}=\xi/a=0.4$, $\bar{R}=R/a=0.6$ and {$\omega_0/\Omega=0.3$}, as function of the rescaled rotation speed $\bar{v}=(\Omega a)/c$ of the detector. For these specific values of parameters, the rate drops to zero for $\bar{v}\lesssim 3.73$. The first threshold corresponds to setting $n=4$ and $\nu=1$ in Eq.~\eqref{solsol}, and it is indicated by a dotted green line in the main panel. Each peak represents the resonant contribution associated to the excitation of one eigenmodes of the cylindrical cavity. In figure we indicate some of the resonance channels that open upon increasing $\bar{v}$, labelled by the pair of quantum numbers $(n,\nu)$.   The inset shows that  the minimal velocity for the onset of spontaneous excitation of the ground state occurs at $n=4$ for $\nu=1$ (green dotted line), in agreement with the main panel.  
 Higher values of $\nu$ yield resonances falling outside the range of $\bar{v}$ plotted  in the main panel (for instance, $\bar{v}_{(n=1,\nu=2)}\simeq 17$ and $\bar{v}_{(n=2,\nu=2)}\simeq9.64$).} 
\label{fig3}
\end{figure}


The transition probability $A_\uparrow$ can be calculated making use of second-order perturbation theory (see~\cite{noteSM}). 
{As it is sketched in Fig.~\ref{fig1},
the} condensate is assumed to be  {radially} confined in a cylinder of radius $a$  {and to extend indefinitely along $z$. The cylindrical confinement is modelled by imposing}  Dirichlet boundary conditions, $\delta\rho(r)|_{r=a}=0$ on the density perturbation (see~\cite{noteSM} and Ref.~\cite{unr} for further details)  {and} implies quantization of the Bogolyubov cylindrical waves, {with spatial mode profiles proportional to  $J_{n} \left( \xi_{n\nu} r/a \right)\, e^{i n \theta}\, e^{i k z}$: for each value $n$ of the angular momentum, the radial momenta $q_{n\nu}\equiv \xi_{n\nu}/a$ are determined by the $\nu$th zero $\xi_{n\nu}$ of the Bessel function $J_n(\cdot)$.}  {Given the infinite size of the BEC along $z$, the linear momentum $k$ can have arbitrary values.}

A plot of the dimensionless ground-state excitation rate, $\Gamma_\uparrow$, as a function of the rescaled detector speed $\bar{v} = \Omega a/c$, is reported in Fig.~\ref{fig3}. {For the Bogolyubov mode of azymuthal and radial quantum numbers $(n,\nu)$,} such excitation rate is non-vanishing only for 
\begin{equation}\label{condition}
n\Omega > q_{n\nu}c\sqrt{1+\left(\frac{q_{n\nu}\xi}{2}\right)^2}+\omega_0,
\end{equation}
{which provides a generalised superradiant condition on the angular velocity to excite a phonon with a given angular momentum $n$. This condition involves the frequency $\omega_0$ of the detector} and the cut-off frequency $\omega_{{n\nu,k=0}}$ of the {branch of} Bogolyubov modes {propagating along the axis of the cylinder in the $n\nu$ {radial-azymuthal} mode{; in our case $\omega_{{n\nu,k=0}}=q_{n\nu}c[1+\left(q_{n\nu}\xi/2\right)^2]^{1/2}$.}

A further trend in the strength of the emission is due to the 
Bessel factor $J_{n} \left( \xi_{n\nu} R/a \right)$ appearing in the mode profile, that suppresses the coupling of the detector to the high angular momentum phonon modes. 
This Bessel factor and, in particular, its strong suppression at short radii $R$ {(for $n>0$)}, puts on rigorous grounds the usual qualitative reasoning based on the local dispersion relation of the waves in the rotating frame, {and on the necessity  of a local super-sonic motion}~\cite{Cal}.


In the geometry under consideration here, $\Gamma_{\uparrow}$ displays peaks whenever the condition~\eqref{condition} holds as an equality {for a given {$n,\nu>0$} pair; this indicates the opening of  a new emission channel occurring at}
\begin{equation}\label{solsol}
\bar{v}_{n,\nu}=\frac{\xi_{n\nu}\sqrt{4+(\xi_{n\nu}y)^2}}{2|{n-\omega_0/\Omega}|}.
\end{equation}
The excitation rate can be then be rewritten as (see~\cite{noteSM})
\begin{equation}
\Gamma_{\uparrow} \propto \sum_{n = 1}^{\infty} \sum_{\nu = 1}^{\infty} \gamma_{n\nu}{(\bar{v})}
  \theta(n-{\omega_0/\Omega}) {\theta(\bar{v}-\bar{v}_{n,\nu})}
\end{equation}
where the square-root divergence
{$\gamma_{n\nu}{(\bar{v})}\propto  [{\bar{v}- \bar{v}_{n,\nu}}]^{1/2}$},
{visible} in Fig.~\ref{fig3}, follows from the effective one-dimensional density of states of {each radial-azymuthal branch of Bogolyubov eigenmodes in the cylindrically-shaped condensate}. Interestingly, the dependence of  $\bar{v}_{n,\nu}$ on $n$ is non-monotonous for a fixed value of $\nu$ as illustrated in the inset of Fig.~\ref{fig3}: this feature explains the non-monotonous labelling of the {peaks visible in the main panel of Fig.~\ref{fig3}.}

{Even though we have restricted our attention to the excitation of the detector, it is useful to recall that this process is always strictly associated with the emission of phonons {propagating away from the detector along the BEC axis $\hat{z}$ (which may  be detected following, for instance, Ref.~\cite{Jeff}).}
In passing, we notice that stimulation of the process by an external incident field, would lead to a superfluid analog of Zeldovich amplification of electromagnetic waves by a rotating dielectric~\cite{zel}.

{As a final remark, it is worth highlighting that the emission processes studied in this work {have  spontaneous nature. They are thus very different from} synchrotron radiation emitted by  circularly moving charges in classical electrodynamics~\cite{Jackson}, in exactly the same way as Ginzburg emission from superluminally moving polarizable objects~\cite{Ginzburg} is conceptually different from the Cherenkov radiation emitted by moving charges or static dipoles~\cite{Cherenkov}. {Analogs of such Cherenkov and synchrotron processes might occur if the  tuning of the Feshbach resonance mentioned in the paragraph before Eq.~\eqref{eq:hami} was not perfect; nevertheless, the statistical properties of the phonon radiation emitted would be drastically different in this case and, more importantly, such processes could not lead to the dynamical instabilities which we  discuss in the following, and which represent one of the salient features of our study.}

\textit{Dynamical instabilities ---} In the astrophysical context, rotational superradiance can give rise to different kinds of dynamical instabilities depending on the specific geometry, from black hole bombs to ergoregion instabilities~\cite{SR:book}. Analog effects are also at play in rotating superfluids~\cite{Giacomelli}. In this final section, we explore dynamical instability mechanisms induced by the circularly moving detector considered in this work. 

{In order to favor self-stimulation of the superradiant process, it is convenient to focus on a {fully confined}, pancake-shaped condensate with discrete Bogolyubov modes. To avoid the instability being disturbed} by saturation of the two-level detector, we {must extend our model by assuming that the internal structure of the detector is well approximated by} a harmonic oscillator coupled to the density fluctuations of the condensate. {To this purpose, one can consider a large number $N$ of two-level atoms whose average distance is smaller than the magnitude of the inverse wave-vector undergoing the dynamical instability: analogously to the Dicke model~\cite{Dicke}, all the atoms can then be  grouped into a large collective spin of size proportional to $N$, with the consequence that saturation effects will start to become relevant only when   an equally large number of  phonons is emitted. Technically, this corresponds to approximate a large spin of size $N$ with harmonic oscillator creation and annihilation operators using an Holstein-Primakoff transformation~\cite{Dicke}, and to introduce a cut-off on the number of excitations in the detector proportional to $N$. }

{Restricting for simplicity our attention to the lowest excitation mode along the $\hat{z}$ direction,} 
the total Hamiltonian in the frame co-moving with the rotating detector is then given by
\begin{equation}
\begin{split}
\label{hamm}
H = \omega_0 d^{\dagger} d 
+\sum_{n,\nu}& \bar{\omega}_{n\nu} 
 \bar{b}^{\dagger}_{n\nu} \bar{b}_{n\nu} 
+ \frac{\mathcal{G}}{\mathcal{N}} x \, \delta\rho({\bf r}_{D}),
\end{split}\end{equation}
%
%
%
%
%
%
{where  $d,d^\dag$ are the harmonic oscillator destruction and creation operators for the impurity and we have defined $x =  (d + d^{\dagger})/\sqrt{2\omega_0}$. Furthermore, we have set $\bar{\omega}_{n\nu} = \omega_{n\nu} - n\Omega$ equal to the Bogolyubov mode frequency in the rotating frame and we indicate with $r_D$ the radial position of the detector.} In practical calculations, the sums will be restricted to $-n_0\leq n\leq n_0$ and $0\leq\nu\leq\nu_0$. For convenience, a factor $\mathcal{ N}\equiv (2n_0+1)\nu_0$ has been included to count the number of cylindrical Bogolyubov modes, so to ensure a proper scaling of the coupling in the multi-mode limit $\mathcal{N}\gg1$.

{In order to identify dynamical instabilities, we consider the corresponding equations of motion,
\bea\label{eq:moto}
\dot{d} &=& -i \left[ \left(\frac{\omega_0}{\Omega}\right)d + \sum_{n = -n_0}^{n_0} \sum_{\nu = 1}^{\nu_0} 
g_{n\nu} ( \bar{b}_{n\nu}  + \bar{b}^{\dagger}_{n\nu} ) \right]
\nn\\
\dot{ \bar{b}}_{n\nu}&=& - i 
\left[ {\tilde{\omega}_{n\nu}}  \bar{b}_{n\nu}
+  g_{n\nu} (d + d^{\dagger}) \right]
\eea
 where we have introduced the dimensionless coupling between modes and impurities 
\begin{equation}
g_{n\nu} \equiv  
\frac{\bar{g}}{{\cal N}} \frac{{\xi_{n\nu}}}{\bar{v}^2}
\frac{{ J_{n} \left( \xi_{n\nu} \bar{R} \right) }}{{ J_{n+1}(\xi_{n\nu}) }} 
\left( \frac{\Omega}{ 2 \omega_0 ( \widetilde{\omega}_{n\nu} + n )  } \right)^{1/2}
\end{equation}
with $\bar{g}=\mathcal{G}\sqrt{\rho_0/m}/c^2$, and $\widetilde{\omega}_{n\nu}=\bar{\omega}_{n\nu}/\Omega$. Furthermore, in the Heisenberg equations~\eqref{eq:moto} time has been rescaled by $\Omega^{-1}$.

%
The different kinds of dynamical instabilities that this Hamiltonian can display are physically understood considering the simplified two-mode bosonic model Hamiltonian
\begin{equation}\label{eqsimp}
H=\omega_0 a^\dag a + \bar{\omega}b^\dag b +g(a^\dag b+a^\dag b^\dag +\textsl{h.c.});
\end{equation}
the eigenvalues of the associated set of  linear Heisenberg equations of motion 
\begin{equation}\label{auto}
\lambda=\pm \frac{i}{\sqrt{2}}\sqrt{\omega^2_0+\bar{\omega}^2-\sqrt{(\omega^2_0-\bar{\omega}^2)^2+16g^2\omega_0\bar{\omega}}}
\end{equation}
allow for direct inspection for the conditions of stability. If $\omega_0$ and $\bar{\omega}$ are both positive, the onset of the first type of dynamical instability occurs for $g>\sqrt{\omega_0\bar{\omega}}/2$, therefore a tiny value of $g$ can induce unstable dynamics in the nearby of $\bar{\omega}\simeq0$. On the other hand, exactly on the parametric resonance $\bar{\omega}+\omega_0=0$, imaginary values of $\lambda$ can be found for any value of $g$. In the  {two-dimensional} many-mode problem defined by~\eqref{hamm}, this latter condition is satisfied
when equation~\eqref{condition} holds as an equality -- a circumstance  {made} possible  {by} the rotational Doppler shift experienced by the Bogolyubov modes.  

 \begin{figure}[t!]
\includegraphics[width=8.6cm]{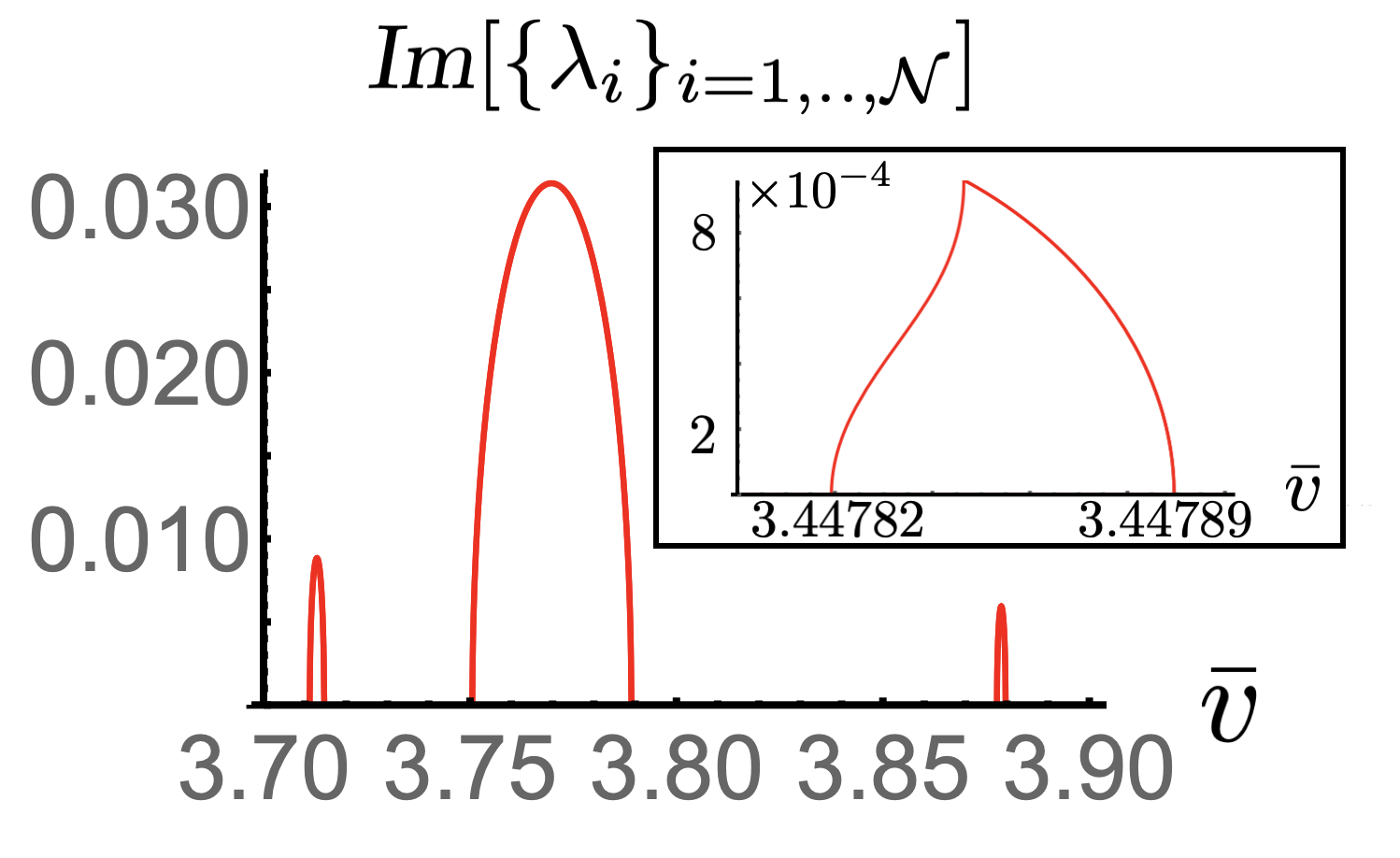}
 \caption{Positive imaginary parts of the eigenvalues of~\eqref{eq:moto} as a function of $\bar{v}$, evaluated for  $\bar{y}=0.4$, $\bar{R}=0.6$, {$\omega_0/\Omega=0.3$}, $\bar{g}/\mathcal{N}\simeq1.53$; a numerical cut-off at $n_0=9$ and $\nu_0=1$  has been used ($\mathcal{N}=19$). The non-vanishing imaginary parts visible in the $\bar{v}$ interval plotted in the figure correspond to the parametric resonances $\bar{\omega}_{n,\nu}+\omega_0=0$ for respectively $n=4,5,6$. The parametric resonance for $n=3$ triggers a small imaginary part of order~$10^{-3}$ (not visible in the figure) around $\bar{v}\simeq3.82$. The dynamical instabilities corresponding to the other peaks of Fig.~\ref{fig3} fall at larger $\bar{v}$ values. Other types of instabilities arise when the conditions $\bar{\omega}_{n,\nu}\simeq0$ are satisfied; the one occurring at $n=3$ and $\nu=1$ is illustrated in the inset. } 
\label{fig5}
\end{figure}

%

This physics is illustrated in Fig.~\ref{fig5} where the imaginary parts of the eigenvalues of the linear system~\eqref{eq:moto} for the fully multi-mode problem are evaluated for the same parameters employed in the plot of Fig.~\ref{fig3}. The resonant condition underlying each instability window is specified in the caption. Different strengths are found for instabilities of the two types. For those of the second kind (akin to parametric instabilities),  a crucial contribution is due to the spatial profile of the mode via the $J_n(\xi_{n\nu}\bar{R})$ factor. Physically, the onset of instabilities {will be observable} as an exponential growth of the amplitude of some Bogolyubov mode {at a rate set by the imaginary part of the eigenvalue}, in concomitance with an {analogous exponential growth of the internal oscillation amplitude of the detector. Even though this instability   mechanism would be quickly saturated for a single two-level impurity after the emission of the first phonon, it can lead to sizeable excitations if several impurities are used to mimic a harmonic oscillator, as discussed at the beginning of this Section.}

{As compared to the notoriously elusive nature of the superradiance effect in the electromagnetic context, we anticipate that the phonon instabilities discussed here can be employed as a way to reinforce the signature of superradiance by inspecting  the quick exponential growth of the detector excitation. Further insight into the underlying process can be provided by the spectral selective and velocity-dependent nature of the amplification mechanism (recall that the unstable modes  are those satisfying the resonance condition {$\bar{\omega}_{n,\nu}\simeq0$}). Experimentally, information on the emitted phonons can in fact be retrieved from the density profile of the BEC and its noise properties.}
{To give a concrete estimate on the timescales required to observe the lasing instability, one can consider the largest real positive eigenvalue $\lambda^*\simeq0.03$ of Fig.~\ref{fig5}, and evaluate the associated time scales in units of $\Omega^{-1}$ (cf. Eqs.~\eqref{eq:moto}), which yields $t^*\simeq 1/(\lambda^*\Omega)\simeq 0.3$s, if $\Omega\simeq100$Hz as in typical experiments for rotating BECs~\cite{cornell}. Since the lifetime of cold atoms   is of the order of many seconds, we expect that the lasing instability discussed in our work should be at reach of state of the art experiments in the field.}


{In addition to their intrinsic interest as a novel manifestation of superradiance, dynamical instabilities triggered by moving detectors are}  of great interest  {also} as a new {concept} of phonon lasing, {where the amplification mechanism is {provided} by the mechanical motion of {a detector} and not by some external pumping}. Even though {self-supported} oscillations are a common feature in classical acoustics {as well as in laser operations},  non-trivial mechanisms for mechanical oscillation accompanied by the onset of quantum fluctuations  {are currently of high interest} in a broad range of platforms, e.g. driven-dissipative coupled microcavities~\cite{jing},  ion-traps~\cite{ions}, nanomagnets~\cite{nano}, and optically driven quantum dots~\cite{opt}.

\textit{Perspectives ---} To summarize, we  have shown in this {Rapid Communication} that a circularly moving {impurity} immersed in an atomic condensate at rest constitutes a promising avenue to investigate {quantum features of rotational superradiance} in a novel context. {If the interaction of the impurity is tuned in a way to serve as a phonon detector, signatures of superradiance} include the excitation of the {impurity} by zero-point quantum fluctuations of the phonon field {in the condensate}, and the onset of dynamical instabilities for the Bogolyubov modes {which can serve as a new avenue for phonon lasing.} 

{Beyond the specific configurations investigated in this work}, our results suggest that moving impurities in condensates can be employed as a novel platform to investigate basic interaction processes between fields in intricate {curved} space-time geometries, with emitters that move at speeds comparable to the wave velocity.
{Different cosmological scenarios and new aspects of trans-Planckian physics with cold atoms can be addressed by tuning the microscopic properties of the BEC, e.g. introducing dipolar interactions as suggested in Ref.~\cite{uuwe}.}
Another intriguing future direction consists of analysing the impact of superradiant effects on higher order quantum vacuum processes such as the Casimir-Polder forces between a pair of circularly rotating impurities. 
{Finally, an exciting challenge is to extend our proposal to photonic quantum simulators, in particular to identify a viable implementation of the moving detector concept in quantum fluids of light~\cite{IC_RMP}.}



 %

 %

\textit{Acknowledgements --- } IC is grateful to Luca Giacomelli and Andrea Vinante for stimulating discussions on the subject of superradiance.

JM is supported by the European Union's Horizon 2020 research and innovation
programme under the Marie Sklodowska-Curie grant agreement No 745608 (QUAKE4PRELIMAT). GM is
supported by Conselho Nacional de Desenvolvimento Cient\'ifico e Tecnol\'ogico -- CNPq under grant 310291/2018-6, and Funda\c{c}\~ao Carlos Chagas Filho de Amparo \`a Pesquisa do Estado do Rio de Janeiro -- FAPERJ under grant E-26/202.725/2018.  IC acknowledges financial support from the Provincia Autonoma di Trento and from the FET-Open Grant MIR-BOSE (737017) and Quantum Flagship Grant PhoQuS (820392) of the European Union.


\newpage

\begin{widetext}

\appendix

\section {\Large Supplemental Material: \\ {Zero-point excitation of a circularly moving detector in an atomic condensate \\ and phonon laser dynamical instabilities}}


\section{\large The model}
\label{model}

As  discussed in Ref.~\cite{Marino2017}, the detector couples to the density fluctuations of a Bose gas via the following Hamiltonian
\beq
H_{I} = g_{-} \sigma_{x}\delta\rho({\bf r}_{A})
\eeq
which  resembles the dipole coupling in quantum electrodynamics. The Hamiltonian that governs the time evolution of the detector is given by
\beq
H_{AA} = \frac{\hbar\omega_0}{2}\sigma_{z}
\eeq
In the above expressions $\sigma_{x,z}$ are the usual Pauli matrices, which can be written in terms of the atomic two-level states $| + \rangle $ (with energy $\hbar\omega_0/2$) and $| - \rangle $ (with energy 
$\hbar\omega_0/2$), namely
\bea
\sigma_{x} &=& |+\rangle\langle-| + |-\rangle\langle+| = \sigma_{+} + \sigma_{-}
\nn\\
\sigma_{z} &=& |+\rangle\langle+| - |-\rangle\langle-|.
\eea
In order to implement our approach, we  switch to  Heisenberg picture. One gets
\beq
H_{AA}(t) = \frac{\hbar\omega_0}{2}\sigma_{z}(t)
\label{h-atom}
\eeq
and
\beq
H_{I}(t) = g_{-}\sigma_{x}(t)\delta\rho(t,{\bf r}_{A}).
\label{h-int}
\eeq
Notice that we have a coupling between atomic operators and the field operator $\delta\rho$ which is effective only on the trajectory of the atoms. Henceforth we employ units such that $\hbar = 1$.

In order to describe the dynamics associated with the scalar field $\delta\rho$ one may resort to the usual Bogoliubov theory of weakly interacting Bose gas~\cite{castin}. Hence one has the following effective Hamiltonian for the scalar field which generates the time evolution with regard to $t$ (neglecting an overall constant term)
\beq
H_{F}(t) = \int d{\bf k}\, \omega_{{\bf k}}\,b^{\dagger}_{{\bf k}}(t)b_{{\bf k}}(t)
\label{h-field}
\eeq
where $b^{\dagger}_{{\bf k}}, b_{{\bf k}}$ are the usual creation and annihilation operators of the scalar field. In addition, $k$ labels the wave vector of the field modes. In the next section we will discuss the field quantization and present explicitly the associated modes. Within the Bogoliubov theory of dilute Bose gas, one has the  dispersion relation (in the laboratory frame where the condensate is at rest)
\beq
\omega_{k} = ck\sqrt{1 + \left(\frac{k\xi}{2}\right)^2},
\label{dispersion}
\eeq
where $\xi = 1/\sqrt{m\mu}$ is the healing length and $c = \sqrt{\mu/m}$ is the local velocity of sound~\cite{books}. The quantity $\mu = \lambda \rho_{0}$ is the chemical potential of the condensate.

The method we employ here consists in identifying two different contributions to the time evolution of an arbitrary atomic observable, namely the vacuum fluctuations and radiation reaction. Afterwards, we rewrite these contributions at a given order in perturbation theory as quantum evolutions given by two effective Hamiltonians and then we compute each of such contributions to the atomic energy level shift. As discussed in Refs.~\cite{cohen2,cohen3}, one should recall that the vacuum-fluctuation term should contain the free part of the field in addition to the contribution of the field to the atomic observable (the source part). In turn, the radiation-reaction contribution should comprise the free part of the atomic observable and, likewise, the source part of the field which emerges due to the presence of the atom itself.  

\section{\large Density correlations}
\label{corr}

Our aim in this section is to evaluate the aforementioned density correlation functions. For a dilute Bose gas, small fluctuations on top of the condensate can be described by the Bogoliubov theory of dilute condensates. The time evolution of the macroscopic wavefunction  is described by the Gross-Pitaevski equation:
\beq
i\frac{\partial}{\partial t}\psi = \left[-\frac{1}{2m} \nabla^{2} + \mu 
+\lambda|\psi|^{2}\right]\psi 
\label{GP}
\eeq
with  chemical potential $\mu$. In the usual formulation to describe elementary excitations in the condensate, one takes a steady state $\psi_{0}$ as the mean field solution,
\beq
\psi(t,{\bf r}) = \psi_{0}({\bf r})[1 + \delta\psi(t,{\bf r})]\,e^{-i \mu t}.
\eeq
From the above equation one may write the atomic Bose gas density as
\beq
\rho(x) = |\psi_{0}({\bf r})|^{2}[1 + \delta\psi(t,{\bf r})+\delta\psi^{\dagger}(t,{\bf r})]
\eeq
where we are keeping only terms that are linear in $\delta\psi$. The operator $\delta\psi$ describing fluctuations satisfies the Bogoliubov-de Gennes equation,
\beq
i\frac{\partial}{\partial t}\delta\psi = - \left[\frac{1}{2m} \nabla^{2} 
+ \frac{1}{m} \frac{\nabla \psi_{0}}{\psi_{0}} \nabla \right] \delta\psi 
+ \lambda \rho_0 \left( \delta\psi + \delta\psi^{\dagger} \right),
\label{bdg}
\eeq
where we assumed an uniform Bose gas, $V_{\textrm{ext}} = 0$. Henceforth we assume that $\rho_0$ is uniform. We search for solutions in cylindrical coordinates of the form 
\beq
\delta\psi = ( \sqrt{\rho_{0}} )^{-1} \sum_{n = -\infty}^{\infty}\frac{1}{2\pi} 
\int_{-\infty}^{\infty} d k \int_{0}^{\infty} dq q
\bigl( u_{qnk}({\bf r}) b_{qnk}(t) + v^{*}_{qnk}({\bf r}) b^{\dagger}_{qnk}(t) \bigr)
\eeq
with commutation relations given by ($s = (qnk)$) 
\bea
[ b_{s},b^{\dagger}_{s^{\prime}} ] &=& \frac{(2\pi)}{q} \delta(q-q^{\prime})\delta(k - k^{\prime})
\delta_{nn^{\prime}}
\nn\\
b_{qnk}(t) &=& b_{qnk}(0) e^{-i \omega_{K} t}
\eea
where $K^2 = q^2 + k^2$ and all other commutators vanish. The mode functions have the form
\bea
u_{qnk}({\bf r}) &=& u_{K} J_{n}(qr) e^{i n \theta} e^{i k z}
\nn\\
v_{qnk}({\bf r}) &=& v_{K} J_{n}(qr) e^{i n \theta} e^{i k z}
\eea 
where $J_{n}(z)$ is the Bessel function of the first kind, with a normalization condition given by
\beq
\int d{\bf r} \left( u^{*}_{s}({\bf r}) u_{s^{\prime}}({\bf r})  
- v^{*}_{s}({\bf r}) v_{s^{\prime}}({\bf r})  \right) = 
\frac{(2\pi)^{2}}{q} \delta(q-q^{\prime})\delta(k - k^{\prime}) \delta_{nn^{\prime}}
\label{normal_cyl}
\eeq
Employing the closure equation for Bessel functions, one finds that
\beq
|u_{K}|^{2} - |v_{K}|^{2} = 1.
\eeq
Using the expression for the Laplacian in cylindrical coordinates, one finds that the dispersion relation is given by
\beq
\omega_{K} = cK\sqrt{1 + \left(\frac{K\xi}{2}\right)^2}.
\eeq 
Without loss of generality, one may consider the coefficients $u_{K}, v_{K}$ as real quantities. One finds 
\bea
u_{K}^{2} &=& \frac{1}{2} \left( \frac{\zeta_{K}}{\omega_{K}} + 1 \right)
\nn\\
v_{K}^{2} &=& \frac{1}{2} \left( \frac{\zeta_{K}}{\omega_{K}} - 1 \right),
\eea
where $\zeta_{K} = E_{K} + \mu$, $E_{K} = K^{2}/2m$. Moreover, $(u_{K} + v_{K})^{2} = 
E_{K}/\omega_{K}$. Hence, the density fluctuations can be written in cylindrical coordinates as
\beq
\delta\rho(t, {\bf r}) = \sqrt{\rho_{0}} \, \sum_{n = -\infty}^{\infty} \frac{1}{2\pi} 
\int_{-\infty}^{\infty} d k \int_{0}^{\infty} dq q (u_{K} + v_{K})
\Bigl( \phi_{qnk} b_{qnk}(t) + \phi_{qnk}^{*} b^{\dagger}_{qnk}(t) \Bigr)
\eeq
where $\phi_{qnk} = J_{n}(qr) e^{i n \theta} e^{i k z}$. As usual, the associated vacuum state is defined as 
$b_{qnk} | 0_{\textrm{M}} \rangle = 0$.

Now let us suppose that the system is confined to a cylinder of radius $a$, on which the fields satisfy Dirichlet boundary conditions. The mode functions are given by
\bea
u_{n \nu k}({\bf r}) &=& u_{n\nu k} J_{n} \left( \frac{\xi_{n\nu} r}{a} \right) e^{i n \theta} e^{i k z}
\nn\\
v_{n \nu k}({\bf r}) &=& v_{n\nu k} J_{n} \left( \frac{\xi_{n\nu} r}{a} \right) e^{i n \theta} e^{i k z}
\eea 
where $\xi_{n\nu}$ is the $\nu$th zero of the Bessel function $J_{n}(x)$. Now the normalization condition reads ($p = (n \nu k)$)
\beq
\int d{\bf r} \left( u^{*}_{p}({\bf r}) u_{p^{\prime}}({\bf r})  
- v^{*}_{p}({\bf r}) v_{p^{\prime}}({\bf r})  \right) = 
(2\pi)^{2} \delta(k - k^{\prime}) \delta_{nn^{\prime}} \delta_{\nu\nu^{\prime}}.
\label{normal_cylb}
\eeq
Using standard relations coming from integrals of Bessel functions, one obtains that
\beq
|u_{n\nu k}|^{2} - |v_{n\nu k}|^{2} = \frac{2}{a^{2} [ J_{n+1}(\xi_{n\nu}) ]^{2}} .
\eeq
Now we find the following dispersion relation
\beq
\omega_{n\nu k} = cK_{n\nu k}\sqrt{1 + \left(\frac{K_{n\nu k} \, \xi}{2}\right)^2}
\eeq 
where $K_{n\nu k}^2  =  \xi_{n\nu}^2/a^2 + k^2$. One also has that
\beq
\delta\psi = ( \sqrt{\rho_{0}} )^{-1} \sum_{n = -\infty}^{\infty} \sum_{\nu = 1}^{\infty}
\frac{1}{2\pi} \int_{-\infty}^{\infty} d k 
\bigl( u_{n\nu k}({\bf r}) b_{n\nu k}(t) + v^{*}_{n\nu k}({\bf r}) b^{\dagger}_{n\nu k}(t) \bigr)
\eeq
and the commutation relations become
\bea
[b_{p},b^{\dagger}_{p^{\prime}}] &=& (2\pi) \delta(k - k^{\prime}) \delta_{nn^{\prime}} \delta_{\nu\nu^{\prime}}
\nn\\
b_{n\nu k}(t) &=& b_{n\nu k}(0) e^{-i \omega_{n\nu k} t}.
\eea
with all other commutators being zero. As above we take the coefficients $u_{n\nu k}, v_{n\nu k}$ to be real. Hence, with an analogous calculation as before, one finds that
\bea
u_{n\nu k}^{2} &=& \frac{1}{2} \left( \frac{\zeta_{n\nu k}}{\omega_{n\nu k}} + 1 \right)
\frac{2}{a^{2} [ J_{n+1}(\xi_{n\nu}) ]^{2}}
\nn\\
v_{n\nu k}^{2} &=& \frac{1}{2} \left( \frac{\zeta_{n\nu k}}{\omega_{n\nu k}} - 1 \right)
\frac{2}{a^{2} [ J_{n+1}(\xi_{n\nu}) ]^{2}} ,
\eea
where $\zeta_{n\nu k} = E_{n\nu k} + \mu$, $E_{n\nu k} = K_{n\nu k}^{2}/2m$. Moreover
$$
(u_{n\nu k} + v_{n\nu k})^{2} = \frac{2}{a^{2} [ J_{n+1}(\xi_{n\nu}) ]^{2}} \frac{E_{n\nu k}}{\omega_{n\nu k}}.
$$ 
Hence, the density fluctuations become
\beq
\delta\rho(t, {\bf r}) = \sqrt{\rho_{0}} \, \sum_{n = -\infty}^{\infty} \sum_{\nu = 1}^{\infty}
\frac{1}{2\pi} \int_{-\infty}^{\infty} d k (u_{n\nu k} + v_{n\nu k})
\Bigl( \psi_{n\nu k} b_{n\nu k}(t) + \psi_{n\nu k}^{*} b^{\dagger}_{n\nu k}(t) \Bigr).
\eeq
where $\psi_{n\nu k} = J_{n} \left( \xi_{n\nu} r/a \right) e^{i n \theta} e^{i k z}$. In this case, the vacuum state is defined as $b_{n\nu k} | 0_{\textrm{C}} \rangle = 0$.\\

A cylindrical coordinate system $(t,r,\bar{\theta},z)$ rigidly rotating at a fixed angular velocity $\Omega$ is related to the cylindrical inertial coordinate system $(t,r,\theta,z)$ by the usual transformation 
$\bar{\theta} = \theta - \Omega t$. In this case, the Bogoliubov-de Gennes equation for a uniform Bose gas (with $\rho_{0}$ uniform) should be properly modified. One finds
\beq
i \left( \frac{\partial}{\partial t} - \Omega \frac{\partial}{\partial \bar{\theta}} \right) \delta\psi 
= -\frac{1}{2m} \nabla^{2} \delta\psi 
+\lambda \rho_0 \left( \delta\psi + \delta\psi^{\dagger} \right),
\label{bdg_rotate}
\eeq
where the Laplacian $\nabla^{2}$ is written in cylindrical coordinates. Now one proceeds with analogous considerations as in the case of the cylindrical inertial coordinate system, with the replacements $\theta \to \bar{\theta}$ and $\omega_{n\nu k} \to \bar{\omega}_{n\nu k}$, where 
$\bar{\omega}_{n\nu k} = \omega_{n\nu k} - n\Omega$, and we are assuming again that the system is confined to a cylindrical mirror of radius $a$. It is evident that the associated rotating modes are not generally of positive frequency. The generalization for this rotating coordinate system is then straightforward. One obtains
\beq
\delta\rho(t, {\bf r}) = \sqrt{\rho_{0}} \,  \sum_{n = -\infty}^{\infty} \sum_{\nu = 1}^{\infty}
\frac{1}{2\pi} \int_{-\infty}^{\infty} d k (u_{n\nu k} + v_{n\nu k})
\Bigl( \bar{\psi}_{n\nu k} \bar{b}_{n\nu k}(t) + \bar{\psi}_{n\nu k}^{*} \bar{b}^{\dagger}_{n\nu k}(t) \Bigr).
\eeq
where $\bar{\psi}_{n\nu k} = J_{n} \left( \frac{\xi_{n\nu} r}{a} \right) e^{i n \bar{\theta}} e^{i k z}$, 
$\bar{b}_{n\nu k}(t) = \bar{b}_{n\nu k}(0) e^{-i \bar{\omega}_{n\nu k} t}$, and
$$
(u_{n\nu k} + v_{n\nu k})^{2} = \frac{2}{a^{2} [ J_{n+1}(\xi_{n\nu}) ]^{2}} 
\frac{E_{n\nu k}}{(\bar{\omega}_{n\nu k} + n\Omega)}
$$ 
with $\bar{\omega}_{n\nu k} + n\Omega > 0$. The creation and annihilation operators $\bar{b}^{\dagger}_{n\nu k}, \bar{b}_{n\nu k}$ satisfy identical commutation relations to those involving canonically conjugated observables in an inertial coordinate system. An analogous situation is discussed in Ref.~\cite{Davies:96}. The authors argue that, due to the fact that the Bogoliubov transformation between the rotating and inertial vacuum states is trivial, the inertial and rotating vacuum should produce identical results. This means that the rotating vacuum $| \bar{0} \rangle$, defined as $\bar{b}_{n\nu k} | \bar{0} \rangle = 0$ should be identical to the vacuum state $| 0_{\textrm{C}} \rangle$ defined above. For more discussion regarding stationary coordinate systems on flat spacetime, we also refer the reader to the Refs.~\cite{Pfautsch:81,Ottewill:03}. \\

Let us present the associated expressions for the correlation function of the density fluctuations. One finds, for the inertial vacuum in an unbounded space
\bea
\langle 0_{M} | \delta\rho(t, {\bf r}) \delta\rho(t', {\bf r}') | 0_{M} \rangle &=&
\frac{\rho_{0}}{ 2 m (2\pi)  } 
\sum_{n = -\infty}^{\infty} \int_{-\infty}^{\infty} d k \int_{0}^{\infty} dq q
\frac{(q^{2} + k^{2})}{\omega_{K}} 
\nn\\
&\times& J_{n}(qr) J_{n}(qr')
e^{i n (\theta-\theta')} e^{i k (z-z')} e^{-i \omega_{K} (t-t')}.
\label{corr_cyl}
\eea
Here the detector's trajectory is then given by $r = R = \textrm{const}$, $z = \textrm{const}$ and $\theta = \Omega t$. On the other hand, if one wishes to switch to a frame co-moving with the rotating detector, one should consider the above results in a rotating coordinate system. This is most easily achieved by performing the calculations in the rotating coordinate system discussed above.  As argued above, in this case we consider that the system is confined inside a cylindrical mirror of radius $a$. One finds that
\bea
\langle \bar{0} | \delta\rho(t, {\bf r}) \delta\rho(t', {\bf r}') | \bar{0} \rangle &=&
\frac{\rho_{0}}{ m (2\pi)  } 
\sum_{n = -\infty}^{\infty} \sum_{\nu = 1}^{\infty} \frac{1}{a^{2} [ J_{n+1}(\xi_{n\nu}) ]^{2}}
\int_{-\infty}^{\infty} d k
\frac{( \xi_{n\nu}^2/a^2 + k^2 )}{(\bar{\omega}_{n\nu k} + n\Omega)} 
\nn\\
&\times& J_{n} \left( \frac{\xi_{n\nu} r}{a} \right) J_{n} \left( \frac{\xi_{n\nu} r'}{a} \right)
e^{i n (\bar{\theta}-\bar{\theta}')} e^{i k (z-z')} e^{-i \bar{\omega}_{n\nu k} (t-t')}
\label{corr_cylb}
\eea
In this case, the detector's trajectory is simply given by $r = R = \textrm{const}$, $z = \textrm{const}$ and $\bar{\theta} = \textrm{const}$. Given the correlation function of the field $\delta\rho$, we can construct the relevant quantities of the formalism employed here. For instance, the normalized symmetric correlation function of the field is given by 
\beq
D(t,t'; {\bf r}, {\bf r}') = \langle 0_{a} |\{ \delta\rho(t,{\bf r}), \delta\rho(t',{\bf r}')\}| 0_{a} \rangle,
\label{hada}
\eeq
where $a = M, C$, depending on the situation that is being under study, whereas the normalized linear susceptibility of the field reads
\beq
\Delta(t,t'; {\bf r}, {\bf r}') =  \langle 0_{a} |[ \delta\rho(t,{\bf r}), \delta\rho(t',{\bf r}')]| 0_{a} \rangle.
\label{pauli}
\eeq
One should consider the free part of the field when evaluating such expressions.


\section{\large Rate of variation of the atomic energy}
\label{ginzburg}

Now one is ready to evaluate the contributions from vacuum fluctuations and radiation reaction to the rate of change of the atom excitation energy. The sum of such contributions produces the total rate. One finds, for large enough $\Delta t = t - t_{0}$ the following expression for the rate of change of the mean atomic excitation energy in the case of the rotating atom coupled with the inertial vacuum state (in an unbounded space)
\beq
\bigg\langle
\frac{d H_{AA}}{dt} \bigg\rangle  
= - g_{-}^2\,
\frac{\rho_{0}}{ 2 m  } 
\sum_{\omega'} \Delta\omega {\cal A}(\omega, \omega') \,
\sum_{n = -\infty}^{\infty} \int_{-\infty}^{\infty} d k \int_{0}^{\infty} dq q
\frac{(q^{2} + k^{2})}{\omega_{K}} 
( J_{n}(qR) )^2
 \delta( \omega_{K} - \Delta\omega - n\Omega).
\eeq
where $\Delta\omega = \omega - \omega'$ and ${\cal A}(\omega, \omega') =  | \langle \omega |\sigma_{x}^{f}(0)| \omega' \rangle |^{2}$, $|\omega\rangle$ and $|\omega'\rangle$ being atomic states (this quantity comes from the atomic correlation functions). The superscript $f$ here means the free part of the atomic operator 
$\sigma_{x}$, as discussed in Sec.~\ref{model}. Observe that, for the rate of change of the atomic excitation energy, when $\Delta \omega > 0$ one has spontaneous emission; otherwise, for $\Delta \omega < 0$ one has spontaneous excitation.

Let us consider that the atom was initially prepared in the state $| + \rangle$. In this case, $\Delta\omega = -\omega_{0}$ and ${\cal A}(\omega, \omega') = 1$. Hence
\beq
\bigg\langle
\frac{d H_{AA}}{dt} \bigg\rangle  
= g_{-}^2\,
\frac{\rho_{0} \omega_{0} }{ 2 m }  
\sum_{n = -\infty}^{\infty} \int_{-\infty}^{\infty} d k \frac{q^{-}_{n}}{\sqrt{k^2 + (q^{-}_{n})^2}}
\frac{F(k,q^{-}_{n})}{G(k,q^{-}_{n})} \left( J_{n}(q^{-}_{n}R) \right)^2
\theta(-\omega_{0} + n\Omega)
\eeq
where 
\bea
F(k,q) &=& \frac{K^3}{\omega_{K}},\,\,\, K = \sqrt{q^2 + k^2}
\nn\\
G(k,q) &=& \frac{c q ( \xi^2 (k^2+q^2)  + 2 ) }{ \sqrt{k^2+q^2} \sqrt{\xi^2 (k^2+q^2) + 4 }}
\nn\\
q^{\pm}_{n} &=& q^{\pm}_{n}(k) = 
\frac{1}{\xi} \left[ \frac{2 \sqrt{c^2+\xi ^2 (\omega_{0} \pm n \Omega )^2}}{c} 
- \left( k^2 \xi^2 + 2\right) \right]^{1/2}.
\eea
Observe that $q^{-}_{n}$ is real (and hence spontaneous excitation may occur) only if
\beq
n \Omega >  k c \sqrt{ 1 + \left(\frac{ k \xi }{2}\right)^2 } + \omega_0
\eeq
where we used the fact that $n$ must be a positive integer. This implies that
\beq
\frac{n \Omega}{c} >  k \sqrt{ 1 + \left(\frac{ k \xi }{2}\right)^2 }.
\eeq

On the other hand, if the atom was initially prepared in the state $| - \rangle$, one would find that 
$\Delta\omega = \omega_{0}$ and ${\cal A}(\omega, \omega') = 1$. Therefore
\beq
\bigg\langle
\frac{d H_{AA}}{dt} \bigg\rangle  
= - g_{-}^2\,
\frac{\rho_{0} \omega_{0} }{ 2 m  }  
\sum_{n = -\infty}^{\infty} \int_{-\infty}^{\infty} d k \frac{q^{+}_{n}}{\sqrt{k^2 + (q^{+}_{n})^2}}
\frac{F(k,q^{+}_{n})}{G(k,q^{+}_{n})}  \left( J_{n}(q^{+}_{n}R) \right)^2
\theta(\omega_{0} + n\Omega).
\eeq
For a positive integer $n$, $q^{+}_{n}$ is real if the following condition is met:
\beq
\omega_{0} >  - n \Omega + k c  \sqrt{1 + \left(\frac{k \xi}{2}\right)^2 }
\eeq
which implies that
\beq
\frac{n \Omega}{c} <  k \sqrt{1 + \left(\frac{k\xi}{2}\right)^2 } .
\eeq

Let us recast our results in terms of dimensionless variables, defined as $X^{\pm}_{n} = q^{\pm}_{n}R$, 
$z = c/(\omega_{0} R)$, $W = \omega_{0}/\Omega$, $\ell(R) = kR$, $y(R) = \xi/R$, and 
$\tilde{v}(R) = \Omega R/c$. Notice that for $\tilde{v}>1$ we are in the supersonic regime. In terms of such variables, the spontaneous excitation rate can be rewritten as 
\beq
\bigg\langle \frac{d H_{AA}}{dt} \bigg\rangle  
= P(\omega_{0}, R) W^{2}  
\sum_{n = -\infty}^{\infty} \int_{-\infty}^{\infty} d \ell \frac{X^{-}_{n}}{\sqrt{\ell^2 + (X^{-}_{n})^2}}
\, \frac{\bar{F}(\ell,X^{-}_{n})}{\bar{G}(\ell,X^{-}_{n})} 
\left( J_{n}(X^{-}_{n}) \right)^2
\theta(-W + n)
\eeq
where 
\beq
P(\omega_{0}, R) = \frac{g_{-}^2 \rho_{0}}{2 m \omega_{0} R^{5} }
\eeq
and
\bea
\omega_{\tilde{v}}(\ell,X^{\pm}_{n}) &=& \frac{ \bigl(\ell^2 + (X^{\pm}_{n})^2 \bigr)^{1/2} }{\tilde{v}}
\sqrt{1 + \left(\frac{\bigl(\ell^2 + (X^{\pm}_{n})^2 \bigr)^{1/2} y}{2}\right)^2}
\nn\\
\bar{F}(\ell,X^{\pm}_{n}) &=&  \frac{\bigl(\ell^2 + (X^{\pm}_{n})^2 \bigr)^{3/2}}
{\omega_{\tilde{v}}(\ell,X^{\pm}_{n})}
\nn\\
\bar{G}(\ell,X^{\pm}_{n}) &=& \frac{X^{\pm}_{n} \Bigl[ y^2 \bigl( \ell^2+(X^{\pm}_{n})^2 \bigr)  + 2 \Bigr] }
{ \tilde{v} \sqrt{\ell^2 + (X^{\pm}_{n})^2} \sqrt{y^2 \bigl(\ell^2 + (X^{\pm}_{n})^2 \bigr) + 4 }}
\nn\\
X^{\pm}_{n} &=& X^{\pm}_{n}(\ell) = \frac{1}{y} \left[ 2 \sqrt{1 + \tilde{v}^2 y^2 (W \pm n)^2} 
- \left( \ell^2 y^2 + 2 \right) \right]^{1/2}.
\eea
For spontaneous emission, one has that
\beq
\bigg\langle \frac{d H_{AA}}{dt} \bigg\rangle  
= -  P(\omega_{0}, R) W^{2}  
\sum_{n = -\infty}^{\infty} \int_{-\infty}^{\infty} d \ell \frac{X^{+}_{n}}{\sqrt{\ell^2 + (X^{+}_{n})^2}}
\, \frac{\bar{F}(\ell,X^{+}_{n})}{\bar{G}(\ell,X^{+}_{n})} 
\left( J_{n}(X^{+}_{n}) \right)^2
\theta(W + n).
\eeq
For clarity we have omitted the dependence on R for the variables $\ell,y,\tilde{v}$ in the above expressions.

Finally, let us consider the rate of variation of the atomic energy with respect to the co-moving frame of the rotating detector. In this situation we confine the system inside a cylindrical mirror of radius $a$. With an almost identical calculation as before, one finds, for a large enough $\Delta t$
\beq
\bigg\langle
\frac{d H_{AA}}{dt} \bigg\rangle  
= - g_{-}^2\,
\frac{\rho_{0}}{ m  } 
\sum_{\omega'}  \,
\sum_{n = -\infty}^{\infty} \sum_{\nu = 1}^{\infty}
 \frac{\Delta\omega {\cal A}(\omega, \omega')}{a^{2} [ J_{n+1}(\xi_{n\nu}) ]^{2}}
\int_{-\infty}^{\infty} d k
\frac{( \xi_{n\nu}^2/a^2 + k^2 )}{(\bar{\omega}_{n\nu k} + n\Omega)} 
\left[ J_{n} \left( \frac{\xi_{n\nu} R}{a} \right) \right]^{2}
 \delta( \omega_{n\nu k} - \Delta\omega - n\Omega).
\eeq
For spontaneous excitation, we find that
\beq
\bigg\langle
\frac{d H_{AA}}{dt} \bigg\rangle  
=  g_{-}^2\,
\frac{2 \rho_{0} \omega_{0} }{ m } 
\sum_{n = -\infty}^{\infty} \sum_{\nu = 1}^{\infty} \frac{ \left[ J_{n} \left( \frac{\xi_{n\nu} R}{a} \right) \right]^{2}}{a^{4} [ J_{n+1}(\xi_{n\nu}) ]^{2}}
\frac{ ( \xi_{n\nu}^2 + ( a k_{0}^{-} )^{2} ) }{\omega_{n\nu k^{-}_{0}}} 
\frac{ \theta(-\omega_{0} + n\Omega) }{{\cal G}_{n\nu k^{-}_{0}}}
\eeq
where
\beq
{\cal G}_{n\nu k} = 
\frac{c k ( \xi ^2 K^{2}_{n\nu k} + 2 ) }{ K_{n\nu k} \sqrt{\xi^2 K^{2}_{n\nu k} + 4 }}
\eeq
and
\beq
k^{\pm}_{0} = \frac{1}{\xi} \left[ \frac{2 \sqrt{c^2+\xi ^2 (\omega_{0} \pm n \Omega )^2}}{c} 
- \left( \frac{\xi^2 \xi_{n\nu}^2}{a^2} +2\right) \right]^{1/2}.
\eeq
Observe that $k^{-}_{0}$ is real (and hence spontaneous excitation may occur) only if
\beq
n \Omega >  \frac{c \, \xi_{n\nu}}{a} \sqrt{ 1 + \left(\frac{ \xi \xi_{n\nu} }{2 a}\right)^2 } + \omega_0
\label{supercond}
\eeq
where we used the fact that $n$ must be a positive integer. This implies  
\beq
\frac{n \Omega a}{c} > \xi_{n\nu} \sqrt{ 1 + \left(\frac{ \xi \xi_{n\nu} }{2 a}\right)^2 }.
\eeq
The lowest bound is obtained for $\nu = 1$. In particular, since the zeros of the Bessel function obey $\xi_{n\nu} > n$, we obtain that the atom remains inert unless $\Omega a > c$. Hence we obtain a similar conclusion as the one in Ref.~\cite{Davies:96}. In other words, if the cylindrical mirror has a radius greater than  $c/\Omega$, then spontaneous excitation can occur. 

In terms of the dimensionless variables $\ell(a) = \bar{x} = k a$, $y(a) = \bar{y} = \xi/a$, 
$\tilde{v}(a) = \bar{v} = \Omega a/c$, $\bar{R} = R/a$ and $X^{2}_{n\nu \bar{x}} = \xi^{2}_{n\nu} 
+ \bar{x}^{2}$, one obtains that
\beq
\bigg\langle
\frac{d H_{AA}}{dt} \bigg\rangle  
=  4 W^{2} P(\omega_{0},a) 
\sum_{n = -\infty}^{\infty} \sum_{\nu = 1}^{\infty} 
 \left[ \frac{J_{n} \left( \xi_{n\nu} \bar{R} \right)}{J_{n+1}(\xi_{n\nu})} \right]^{2}
\frac{X^{2}_{n\nu \bar{x}^{-}_{0}}}{\tilde{\omega}_{n\nu \bar{x}^{-}_{0}}} 
\frac{ \theta(-W + n) }{{\cal G}_{n\nu \bar{x}^{-}_{0}}(\bar{v})}
\eeq
where
\bea
\tilde{\omega}_{n\nu \bar{x}} &=& \frac{X_{n\nu \bar{x}}}{\bar{v}}
\sqrt{1 + \left(\frac{X_{n\nu \bar{x}} \, \bar{y}}{2}\right)^2}
\nn\\
{\cal G}_{n\nu \bar{x}}(\bar{v}) &=& 
\frac{\bar{x} ( \bar{y}^2 X^{2}_{n\nu \bar{x}} + 2 )}{ \bar{v} X_{n\nu \bar{x}} 
\sqrt{  \bar{y}^2 X^{2}_{n\nu \bar{x}} + 4 }}
\nn\\
\bar{x}^{\pm}_{0} &=& \frac{1}{\bar{y}} \left[ 2 \sqrt{1 + \bar{v}^2\bar{y}^2 (W \pm n)^2} 
- \left( \bar{y}^2 \xi_{n\nu}^2 + 2 \right) \right]^{1/2}.
\eea 
On the other hand, for spontaneous emission, we find that
\beq
\bigg\langle
\frac{d H_{AA}}{dt} \bigg\rangle  
=  - g_{-}^2\,
\frac{2\rho_{0} \omega_{0} }{ m } 
\sum_{n = -\infty}^{\infty} \sum_{\nu = 1}^{\infty} \frac{ \left[ J_{n} \left( \frac{\xi_{n\nu} R}{a} \right) \right]^{2}}{a^{4} [ J_{n+1}(\xi_{n\nu}) ]^{2}}
\frac{( \xi_{n\nu}^2 + (a k^{+}_{0})^2 )}{\omega_{n\nu k^{+}_{0}}} 
\frac{ \theta(\omega_{0} + n\Omega) }{{\cal G}_{n\nu k^{+}_{0}}}.
\eeq
For a negative integer $n$, the condition for $k^{+}_{0}$ to be real is similar to the one derived above for $k^{-}_{0}$. On the other hand, for a positive integer $n$, $k^{+}_{0}$ is real if the following condition is met:
\beq
\omega_{0} >  - n \Omega + \frac{c \, \xi_{n\nu}}{a} \sqrt{1 + \left(\frac{\xi \xi_{n\nu}}{2 a}\right)^2 }
\eeq
which implies that
\beq
\frac{n \Omega a}{c} <  \xi_{n\nu} \sqrt{1 + \left(\frac{\xi \xi_{n\nu}}{2 a}\right)^2 } .
\eeq
In terms of the dimensionless variables defined above, one has that
\beq
\bigg\langle
\frac{d H_{AA}}{dt} \bigg\rangle  
=  - 4 W^{2} P(\omega_{0},a) 
\sum_{n = -\infty}^{\infty} \sum_{\nu = 1}^{\infty} 
 \left[ \frac{J_{n} \left( \xi_{n\nu} \bar{R} \right)}{J_{n+1}(\xi_{n\nu})} \right]^{2}
\frac{X^{2}_{n\nu \bar{x}^{+}_{0}}}{\tilde{\omega}_{n\nu \bar{x}^{+}_{0}}} 
\frac{ \theta(W + n) }{{\cal G}_{n\nu \bar{x}^{+}_{0}}(\bar{v})} .
\eeq

\section{\large Einstein coefficients}

It is not difficult to display the Einstein coefficients for spontaneous emission, denoted by $A_{\downarrow}$, and spontaneous excitation, given by $A_{\uparrow}$. Using a similar approach as employed in Ref.~\cite{Audretsch:94}, one finds
\beq
A_{\uparrow} = P(\omega_{0}, R) \frac{\omega_{0}}{\Omega^{2}}  
\sum_{n = 1}^{\infty} \int_{-\infty}^{\infty} d \ell \frac{X^{-}_{n}}{\sqrt{\ell^2 + (X^{-}_{n})^2}}
\, \frac{\bar{F}(\ell,X^{-}_{n})}{\bar{G}(\ell,X^{-}_{n})} 
\left( J_{n}(X^{-}_{n}) \right)^2
\theta(-W + n)
\label{Einstein_up}
\eeq
and
\beq
A_{\downarrow} = P(\omega_{0}, R) \frac{\omega_{0}}{\Omega^{2}}   
\sum_{n = -\infty}^{\infty} \int_{-\infty}^{\infty} d \ell \frac{X^{+}_{n}}{\sqrt{\ell^2 + (X^{+}_{n})^2}}
\, \frac{\bar{F}(\ell,X^{+}_{n})}{\bar{G}(\ell,X^{+}_{n})} 
\left( J_{n}(X^{+}_{n}) \right)^2
\theta(W + n).
\label{Einstein_down}
\eeq
One can define the following dimensionless Einstein coefficients from the above expressions:
\beq
\Gamma_{\uparrow} = \frac{\omega_{0}}{ P(\omega_{0}, R) } \, A_{\uparrow} =   
W^{2} \sum_{n = 1}^{\infty} \int_{-\infty}^{\infty} d \ell \frac{X^{-}_{n}}{\sqrt{\ell^2 + (X^{-}_{n})^2}}
\, \frac{\bar{F}(\ell,X^{-}_{n})}{\bar{G}(\ell,X^{-}_{n})} 
\left( J_{n}(X^{-}_{n}) \right)^2
\theta(-W + n)
\label{Einstein_updim}
\eeq
and
\beq
\Gamma_{\downarrow} = \frac{\omega_{0}}{ P(\omega_{0}, R) } \, A_{\downarrow} =   
W^{2} \sum_{n = -\infty}^{\infty} \int_{-\infty}^{\infty} d \ell \frac{X^{+}_{n}}{\sqrt{\ell^2 + (X^{+}_{n})^2}}
\, \frac{\bar{F}(\ell,X^{+}_{n})}{\bar{G}(\ell,X^{+}_{n})} 
\left( J_{n}(X^{+}_{n}) \right)^2
\theta(W + n).
\label{Einstein_downdim}
\eeq
Defining the excitation rate per mode as $d \Gamma/ d\ell$, one finds, for large $\tilde{v}$ (at leading order)
\beq
\frac{d \Gamma_{\uparrow}}{d\ell} \approx \tilde{v}^{3/2} {\cal F}(\tilde{v})
\eeq
where
\bea
{\cal F}(\tilde{v}) &=& 4W^{2} \sum_{n = 1}^{\infty}
\frac{\theta(-W + n)}{ \pi  \bigl( 2(-W+n) y^3 \bigr)^{1/2}}
\cos^{2} \left( \frac{{\cal B}^{-}}{4}  \right)  
\nn\\
{\cal B}^{\pm} &=& {\cal B}^{\pm}(\ell, \tilde{v}) =
\frac{\sqrt{2} \left(\ell^2 y^2+2\right)}{\sqrt{(\pm W+n) \tilde{v} y^3}}
- 4 \sqrt{2} \sqrt{\frac{(\pm W+n) \tilde{v}}{y}} + 2 \pi n + \pi. 
\eea
Finally, for the case of the rotating vacuum (with the system confined inside a cylindrical mirror), one finds that
\beq
A_{\uparrow} = \frac{4 \omega_{0}}{\Omega^{2}} P(\omega_{0},a) 
\sum_{n = 1}^{\infty} \sum_{\nu = 1}^{\infty} 
 \left[ \frac{J_{n} \left( \xi_{n\nu} \bar{R} \right)}{J_{n+1}(\xi_{n\nu})} \right]^{2}
\frac{X^{2}_{n\nu \bar{x}^{-}_{0}}}{\tilde{\omega}_{n\nu \bar{x}^{-}_{0}}} 
\frac{ \theta(-W + n) }{{\cal G}_{n\nu \bar{x}^{-}_{0}}(\bar{v})}
\label{Einstein2_up}
\eeq
and
\beq
A_{\downarrow} = \frac{4\omega_{0}}{\Omega^{2} } P(\omega_{0},a) 
\sum_{n = -\infty}^{\infty} \sum_{\nu = 1}^{\infty} 
 \left[ \frac{J_{n} \left( \xi_{n\nu} \bar{R} \right)}{J_{n+1}(\xi_{n\nu})} \right]^{2}
\frac{X^{2}_{n\nu \bar{x}^{+}_{0}}}{\tilde{\omega}_{n\nu \bar{x}^{+}_{0}}} 
\frac{ \theta(W + n) }{{\cal G}_{n\nu \bar{x}^{+}_{0}}(\bar{v})} .
\label{Einstein2_down}
\eeq
Accordingly, the associated dimensionless coefficients are given by
\beq
\Gamma_{\uparrow} = \frac{\omega_{0}}{ P(\omega_{0}, a) } \, A_{\uparrow} =
4 W^{2} \sum_{n = 1}^{\infty} \sum_{\nu = 1}^{\infty} 
 \left[ \frac{J_{n} \left( \xi_{n\nu} \bar{R} \right)}{J_{n+1}(\xi_{n\nu})} \right]^{2}
\frac{X^{2}_{n\nu \bar{x}^{-}_{0}}}{\tilde{\omega}_{n\nu \bar{x}^{-}_{0}}} 
\frac{ \theta(-W + n) }{{\cal G}_{n\nu \bar{x}^{-}_{0}}(\bar{v})}
\label{Einstein2_updim}
\eeq
and
\beq
\Gamma_{\downarrow} = \frac{\omega_{0}}{ P(\omega_{0}, a) } \, A_{\downarrow} = 
4 W^{2} \sum_{n = -\infty}^{\infty} \sum_{\nu = 1}^{\infty} 
 \left[ \frac{J_{n} \left( \xi_{n\nu} \bar{R} \right)}{J_{n+1}(\xi_{n\nu})} \right]^{2}
\frac{X^{2}_{n\nu \bar{x}^{+}_{0}}}{\tilde{\omega}_{n\nu \bar{x}^{+}_{0}}} 
\frac{ \theta(W + n) }{{\cal G}_{n\nu \bar{x}^{+}_{0}}(\bar{v})} .
\label{Einstein2_downdim}
\eeq
For large $\bar{v}$ one obtains, at leading order
\beq
\Gamma_{\uparrow} \approx \bar{v}^{3/2} {\cal C}_{n\nu}^{-}
\eeq
where
\beq
{\cal C}_{n\nu}^{\pm} =
8 W^{2} \sum_{n = 1}^{\infty} \sum_{\nu = 1}^{\infty} 
 \left[ \frac{J_{n} \left( \xi_{n\nu} \bar{R} \right)}{J_{n+1}(\xi_{n\nu})} \right]^{2}
 \frac{\theta(\pm W + n)}{\sqrt{2( \pm W+n) y^{3}}} .
\eeq

\end{widetext}

\end{document}